\documentclass[aps,preprint,showpacs]{revtex4-1}
\usepackage{amsfonts}
\usepackage{amsmath}
\usepackage{amssymb}
\usepackage{graphicx}
\usepackage{rotating}
\usepackage{hyperref}
\usepackage{color}   %May be necessary if you want to color links
\usepackage[dvipsnames]{xcolor}

\begin{document}

%\centerline{PA v1.0 23/9/2014; NNS v2.0 24/9/2014,NNS v3.0
%12/12/2014,  v10 NNS 28/1/2015}

\title{Cold magnetized quark matter phase diagram within
a generalized SU(2) NJL model}

\author{P. G. Allen$^a$, V. Pagura$^{b,c}$ and N. N. Scoccola$^{a,b,d}$ }
\affiliation{
$^a$ Department of Theoretical Physics, Comisi\'on
Nacional de Energ\'ia At\'omica, Av.Libertador 8250, 1429 Buenos Aires, Argentina\\
$^b$CONICET, Rivadavia 1917, 1033 Buenos Aires, Argentina\\
$^{c}$IFLP, CONICET $-$ Dpto.\ de F\'{\i}sica, Universidad Nacional de La
Plata, C.C. 67, (1900) La Plata, Argentina\\
$^d$ Universidad Favaloro, Sol\ 453, 1078 Buenos Aires, Argentina}


\begin{abstract}

We study the effect of intense magnetic fields on the phase
diagram of cold, strongly interacting matter within an extended
version of the Nambu-Jona-Lasinio model that includes flavor
mixing effects and vector interactions. Different values of the
relevant model parameters in acceptable ranges are considered.
Charge neutrality and beta equilibrium effects, which are
specially relevant to the study of compact stars, are also taken
into account. In this case the behavior of leptons is discussed.

\end{abstract}

\maketitle

\section{Introduction}
The influence of intense magnetic fields on the properties of
strongly interacting matter has become an issue of increasing
interest in recent years~\cite{1}. This is mostly motivated by the
realization that in some relevant physical situations, like high
energy non-central heavy ion collisions~\cite{2} and compact
stellar objects called magnetars~\cite{3}, very strong magnetic
fields may be produced. Since in these systems extreme
temperatures and/or densities may be found, it is interesting to
investigate which modifications are induced by the presence of
strong magnetic fields on the whole QCD phase diagram.
Unfortunately, even in the absence of those fields, the present
knowledge of such phase diagram is only schematic due to the
well-known difficulty given by the so-called sign problem which
affects lattice calculations at finite chemical
potential~\cite{5}. Of course, the presence of strong magnetic
fields makes the situation even more complex. Thus, most of our
present knowledge of their effect comes from investigations
performed in the framework of effective models (see e.g.~\cite{6}
and refs. therein). In this contribution we present some results
of a study of the phase diagram of cold quark matter subject to
intense magnetic fields in the framework of a generalized
Nambu-Jona-Lasinio (NJL) model. The NJL-type models are effective
relativistic quark models for non perturbative QCD, where gluon
degrees of freedom are integrated out and interactions are
modelled through point like interactions. In its simplest
version~\cite{njl} it only includes scalar and pseudo scalar
interactions that describe chiral symmetry breaking effectively.
As well known, however, a more detailed description of the
low-energy quark dynamics requires that other channels like flavor
mixing and vector meson interactions are taken into
account~\cite{reports}. In fact, some aspects of the effect of
those interactions on the magnetized quark matter have already
been investigated~\cite{Boomsma:2009yk,Denke:2013gha}. The purpose
of the present work is to extend those analyzes by performing a
detailed study of the resulting cold matter phase diagrams,
including their dependence on the parameters that regulate the
strength of these interactions. Moreover, the behavior of cold
magnetized quark matter under conditions relevant for the physics
of compact stars will also be considered. One of the phenomena to
be discussed in detail is that related to the so-called inverse
magnetic catalysis (IMC) expected to exist at low temperature and
moderate values of the chemical potentials ~\cite{Preis:2010cq}.
It is important to mention that, despite bearing a similar name,
this phenomenon is different from the inverse magnetic catalysis
at finite temperature found in lattice QCD (LQCD). Concerning the
latter one, we recall that most effective models foresee that at
zero chemical potential a crossover transition is obtained at a
pseudo critical temperature $T_c$ that increases with an
increasing magnetic field, a behavior which is contrary to the one
found in LQCD calculations \cite{Bruckmann:2013oba}. In fact,
recently there have been significant efforts to modify the models
such that they incorporate a mechanism that could lead to inverse
magnetic catalysis around $T_c$ (see Ref.~\cite{Andersen:2014xxa}
for a recent review on this issue). We should stress, however,
that this is not expected to affect the low temperature behavior
discussed in this work.

This paper is organized as follows. In Sec.~II we provide some
details of the model and its parametrizations as well as the way
to deal with an external constant magnetic field. In Sec.~III we
present and discuss our results for symmetric quark matter. The
situation for stellar matter is analyzed in Sec.~IV. Finally, our
conclusions are given in Sec.~V.

 \section{Formalism}

We consider a generalized NJL-type SU(2) Lagrangian density which
includes a scalar-pseudoscalar interaction, vector-axial vector
and the t\'{}Hooft determinant interaction~\cite{reports}. In the
presence of an external magnetic field and chemical potential it
reads:
\begin{eqnarray}
\mathcal{L}=\bar{\psi}\left( i\ \rlap/\!D   - m_c + \hat{\mu}\
\gamma^0 \right)\psi+ \mathcal{L}_{int}
\end{eqnarray}
where
\begin{eqnarray}
\mathcal{L}_{int} &=& G_1 \sum_{a=0}^{3} \left[ \left(
\bar{\psi}\tau_{a}\psi \right)^{2}+ \left(
\bar{\psi}i\gamma_{5}\tau_{a}\psi \right)^{2} \right] + G_2
\sum_{a=0}^{3} \left[ \left( \bar{\psi}\gamma_\mu \tau_{a}\psi
\right)^{2}+ \left( \bar{\psi}\gamma_\mu\gamma_{5}\tau_{a}\psi
\right)^{2} \right]
\nonumber \\
& & + G_3 \left[ \left( \bar{\psi}\gamma_\mu \psi \right)^{2}+
\left( \bar{\psi}\gamma_\mu\gamma_{5}\psi \right)^{2} \right] +
G_4 \left[ \left( \bar{\psi}\gamma_\mu \psi \right)^{2}- \left(
\bar{\psi}\gamma_\mu\gamma_{5}\psi \right)^{2} \right] + 2 G_{D}
\left( d_{+}+d_{-}\right) \label{lf}
\end{eqnarray}
Here, $G_i$ with $i=1,4$ and $G_D$ are coupling constants, $\psi
=\left(u,d\right)^{T}$ represents a quark field with two flavors,
$d_{\pm}=det\left[ \bar{\psi}\left( 1\pm\gamma_{5} \right)\psi
\right]$, $\hat{\mu}=\mathrm{diag}\left(\mu_{u},\mu_{d}\right)$
the quark chemical potentials, $m_c$ is the (current) mass matrix
that we take to be the same for both flavors, $\tau_{0}= I$, where
$I$ is the unit matrix in the two flavor space, and $\tau_{a}$,
$0~<~a~\leq~3$ denote the Pauli matrices. The coupling of the
quarks to the electromagnetic field ${\cal A}_\mu$ is implemented
through the covariant derivative $D_{\mu}=\partial_\mu - i \hat q
{\cal A}_{\mu}$ where $\hat q$ represents the quark electric
charge matrix $\hat q =\mathrm{diag}\left( q_{u},q_{d} \right)$
where $q_u/2 = - q_d = e/3$. In the present work we consider a
static and constant magnetic field in the 3-direction, ${\cal
A}_\mu=\delta_{\mu 2} x_1 B$. In the mean-field approximation the
associated grand-canonical thermodynamical potential for cold and
dense quark matter reads
\begin{equation}\label{lf_mfa}
  \Omega^{\mathrm{MFA}} = - \sum_{f=u,d} \theta_f + \Omega_{pot}
\end{equation}
where $\theta_{f}$ gives the contribution from the gas of
quasi-particles of each flavor $f=u,d$ and can be written as the
sum of 3 contributions~\cite{Menezes:2008qt}
\begin{eqnarray}
\theta^{\mathrm{vac}}_{f} &=& \frac{N_c}{8\pi^2} \ \left \{
\Lambda\left(2 \Lambda^2 +  M_f^2 \right )\sqrt{\Lambda^2+M_f^2} -
M_f^4 \ln \left[\frac{(\Lambda+ \sqrt{\Lambda^2+M_f^2})}{M_f}
\right]\right \} \mbox{,}
\nonumber \\[2.mm]
\theta^{\mathrm{mag}}_{f} &=& \frac {N_{c} }{2 \pi^2} \
(|q_{f}|B)^2 \ \left [ \zeta^{(1,0)}(-1,x_{f}) -  \frac {1}{2}(
x_{f}^{2} - x_{f}) \ln x_{f} +\frac {x_{f}^{2}}{4} \right ]
\mbox{,}
\nonumber \\[2.mm]
\theta^{\mathrm{med}}_{f}&=& \frac{N_{c}}{4\pi^{2}} \ |q_{f}|B \
\sum_{\nu=0}^{\nu^{max}_f} \alpha_\nu \left[ \tilde \mu_f
\sqrt{\tilde\mu_f^2-s_{f}(\nu,B)^{2}}
\right.\nonumber\\
& & \left. \qquad \qquad \qquad -s_{f}(\nu,B)^{2}\ln\left(
\frac{\tilde\mu_f+\sqrt{\tilde\mu_f^2-s_{f}(\nu,B)^{2}}}{s_{f}(\nu,B)}
\right) \right]  \mbox{,} \label{PmuB}
\end{eqnarray}

\noindent where $M_f = m_c + \sigma_f$ and $\tilde \mu_f = \mu_f -
\omega_f$, with $\sigma_f$ and $\omega_f$ being the mean field
values of the scalar and vector meson fields, respectively.
$\Lambda$ represents a non covariant ultraviolet cutoff and
$\zeta^{(1,0)}(-1,x_f)= d\zeta(z,x_f)/dz|_{z=-1}$ where
$\zeta(z,x_f)$ is the Riemann-Hurwitz zeta function. In addition,
$s_{f}(\nu,B)=\sqrt{M_{f}^2+2|q_{f}|B\nu}$ while $x_f =
M_{f}^{2}/(2 |q_{f}| B)$. In $\theta^{med}_f$, the sum is over the
Landau levels (LL´s), represented by $\nu$, while
$\alpha_\nu=2-\delta_{\nu 0}$ is a degeneracy factor and
$\nu^{max}_f$ is the largest integer that satisfies $\nu^{max}_f
\le (\tilde\mu_f^2 - M^2_f)/(2 |q_f| B)$.

The $\Omega_{pot}$ contribution reads
\begin{eqnarray}
\Omega_{pot} =
      \frac{ (1-c_s) (\sigma_u^2 + \sigma_d^2) - 2 c_s \, \sigma_u \, \sigma_d }{8 g_s ( 1 - 2 c_s)}
      - \frac{ (1-2 c_v) (\omega_u^2 + \omega_d^2) + c_v \, \omega_u \, \omega_d }{8 g_v ( 1 - 2 c_v)}
      \label{Omepot}
\end{eqnarray}
where we have introduced a convenient parametrization of the
coupling constants in terms of the quantities $g_s$, $c_s$, $g_v$
and $c_v$ given by
\begin{eqnarray}
&&g_s = G_1 + G_D \qquad ;  \qquad g_v = G_2 + G_3 + G_4 \nonumber \\
&&c_s=\frac{G_D}{G_1 + G_D} \qquad ; \qquad c_v=\frac{G_3 +
G_4}{2(G_2 + G_3 + G_4)}
\end{eqnarray}

The relevant gap equations are given by
\begin{eqnarray}
\frac{
\partial \Omega^{\mathrm{MFA}}(\sigma_u,\sigma_d,\omega_u,\omega_d)
} {\partial(\sigma_u,\sigma_d,\omega_u,\omega_d)} = 0 \ .
\label{gapeq}
\end{eqnarray}

The solutions to the gap equations minimize the thermodynamic
potential with respect to the quark masses $M_f$, but the
$\omega_f$ derivatives amount to a consistency condition and the
potential is actually a maximum with respect to these variables.
Several solutions will generally exist, corresponding to different
possible phases, and the most stable solution is that which
minimizes the thermodynamic potential with respect to $M_f$.

In our calculations we will consider first the simpler case of
symmetric matter where both quarks carry the same chemical
potential $\mu$. Afterwards, we will analyze the case of stellar
matter in which leptons are also present and $\beta$-equilibrium
and charge neutrality are imposed. In this case the chemical
potential for each quark, $\mu_f$, is a function of quark number
chemical potential $\mu$ and the lepton chemical potentials which
have to be self-consistently determined.

In order to analyze the dependence of the results on the model
parameters, we will consider two SU(2) NJL model
parameterizations. Set~1 corresponds to that leading to
$M_0=340$~MeV while Set~2 to that leading to $M_0=400$~MeV. Here,
$M_0$ represents the vacuum quark effective mass in the absence of
external magnetic fields. The corresponding model parameters are
listed in Table~\ref{pnjl}.

\begin{table}[h]
\caption{\label{pnjl} Parameter sets for the NJL SU(2) model.}
\begin{center}
\begin{tabular}{cccccc}
  \hline  \hline
\hspace*{.2cm} Parameter set \hspace*{.2cm} & \hspace*{.2cm} $M_0$
\hspace*{.2cm}& \hspace*{.2cm} $m$ \hspace*{.2cm} & \hspace*{.2cm}
$g_s \Lambda^2$ \hspace*{.2cm}    & \hspace*{.2cm} $\Lambda$
\hspace*{.2cm}    &
\hspace*{.2cm} $-<u\bar u>^{1/3}$ \hspace*{.2cm}  \\
         &    MeV    &   MeV    &         &  MeV   &  MeV  \\ \hline
Set 1    &    340    &  5.595   & 2.212   &  620.9 & 244.3 \\
Set 2    &    400    &  5.833   & 2.440   &  587.9 & 240.9
\\
\hline
\end{tabular}
\end{center}
\end{table}

The presence of the t'Hooft determinant interaction is very
important since it reflects the $U_A(1)-$anomaly of QCD. Its
strength, and consequently the amount of flavor mixing induced by
this term, is controlled by the parameter $c_s$. An estimate for
its value can be obtained from the $\eta - \eta '$ mass splitting
within the $3-$flavor NJL model~\cite{Kunihiro:1989my}. This leads
to $c_s\simeq 0.2$~\cite{Frank:2003ve}. In any case, to obtain a
full understanding of the effects of flavor mixing we will vary
the value of $c_s$ in a range going from $0$, which corresponds to
a situation in which the two flavors are completely decoupled, to
$0.5$, being this the case of maximum flavor mixing described for
example in Ref.~\cite{Allen:2013lda}. Regarding the vector
coupling term it is important to recall that one can obtain
naturally the terms proportional to $G_2$ if one starts from a
QCD-inspired color current-current interaction and then performs a
Fierz transform into color-singlet channels, and that in this case
the relation between coupling strengths is
$G_2=G_1/2$~\cite{reports}. Yet, the value of $g_v$ cannot be
accurately determined from experiments nor from lattice QCD
simulations and this is why it has been taken as a free parameter
in most works. In the present work we take $0<g_v/g_s < 0.5$. It
is worth mentioning that due to the mixing of pseudoscalar and
longitudinal axial vector interaction terms, pseudoscalar meson
properties depend on $G_2$. Thus, strictly speaking the parameters
given in Table 1 only lead to the empirical values of $f_\pi$ and
$m_\pi$ when $G_2 = 0$. However, as shown in
Ref.~\cite{Hanauske:2001nc}, $m_{\pi}$ and $f_{\pi}$ only change
by $\sim 10\%$ when $G_2/G_1$ increases from $0$ to $1$. Thus, for
simplicity, we will keep the model parameter values fixed when
varying $g_v$. A last comment regarding $c_v$, i.e. the parameter
that regulates the ratio between the singlet and octet
vector-axial vector interaction strengths: we will take it as a
free parameter in the range $0 \leq c_v \leq 1/2$. Note that for
$c_v = 1/2$ only singlet vector-axial interactions are present
and, thus, there is no mixing between the pseudoscalar and
longitudinal axial vector channels.

We end this section by describing the way in which the different
phases of the magnetized quark matter will be denoted as well as
the procedure used to identify the boundaries between them. For
the phases we adopt the notation of
Refs.~\cite{Allen:2013lda,ebert}. Thus, the vacuum (i.e. fully
chirally broken) phase is denoted by B, the massive phases in
which $M_f$ depends on the chemical potential by $\mbox{C}_\alpha$
and, finally, the chirally restored phases by $\mbox{A}_{\alpha}$.
Here, $\alpha$ is a set of two numbers indicating the highest LL
populated for each flavor.  To obtain the critical chemical
potentials at a given $eB$ we proceed as follows. In the case of
first order phase transitions we calculate the thermodynamical
potential  for each of the neighboring phases (that is, the global
minimum with respect to $M_f$) and then search for the chemical
potential values at which they become degenerate. In the case of
crossover transitions the critical value is identified by the peak
of the chiral susceptibility corresponding to each quark flavor,
defined as $\partial{<\overline{\psi}\psi>}/\partial{m_f} $.

\section{Results for Symmetric matter}

In this section we present the results obtained for the case of
symmetric matter. These results were obtained solving the set of
coupled ``gap equations'' (\ref{gapeqs}) for different values of
magnetic field and chemical potential.

\subsection{Effect of the flavor mixing interactions}

To neglect vector interactions implies taking $G_2 = G_3 = G_4 =
0$ in Eq.~(\ref{lf}), while $G_1 \neq 0$ and $G_D \neq 0$. Therefore,
Eqs.~(\ref{gapeq}), will become a set of two coupled equations that must be
solved for the independent variables $M_u$ and $M_d$. The
parameter $c_s$ acts as a coupling between both flavors and we
study how the phase transitions are modified as we vary this
parameter in the range $0 < c_s < 0.5$. The case $c_s = 0.5$
corresponds to ordinary NJL where flavor mixing is maximum and,
thus, both flavors have identical behavior. In fact, the first
term in Eq.~(\ref{Omepot}) will tend to infinity as $c_s$ goes to $0.5$
unless $M_u = M_d$, which leaves only one equation to be solved.
But if $c_s < 0.5$,  then both masses will be independent
variables, and transitions for each flavor might occur
simultaneously or not in different regions of the phase diagrams.

\begin{figure}
\begin{center}
\includegraphics[width=0.8\linewidth,angle=0]{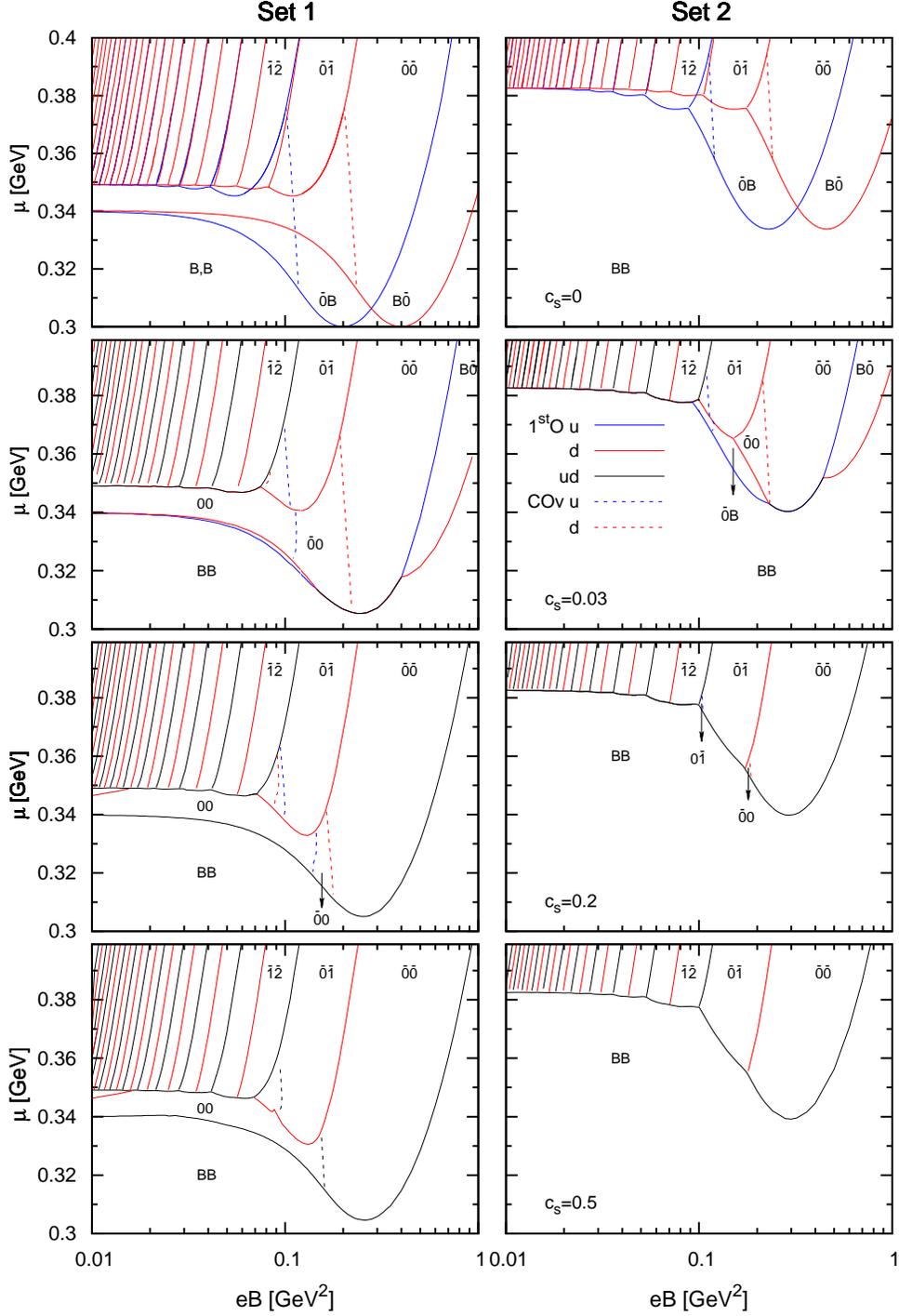}
\end{center}
\caption{\label{fig1}(Color online) Phase diagrams in the $eB-\mu$
plane for different values of flavor-mixing parameter $c_s$. To simplify
the figure we have introduced a compact notation to indicate the
phases. The pair of integers $mn$ corresponds to the
$\mbox{C}_{mn}$ phase and the pair ${\bar m} {\bar n}$ to the
$\mbox{A}_{mn}$ phase. The case in which one quark is in a C-type
phase and the other in the A-type phase is indicated by putting a
bar on top of the integer associated with the A-type phase.}
\end{figure}

The phase diagrams for both parameter sets and several values of
flavor mixing can be seen in Fig.~\ref{fig1}. We will start by
commenting some general features which are common to all phase
diagrams discussed in this work. It is seen that chiral symmetry
is completely broken for chemical potentials well below $M_0$ and
that restoration occurs for high enough chemical potentials,
usually accompanied by a large drop in the dressed mass. The
inclusion of a constant magnetic field modifies the quark
dispersion relation, introducing Landau levels (LL's) into its
spectrum. A consequence of this is that chiral symmetry
restoration might occur in several steps as chemical potential is
increased, each of which is a transition where quark population
appears on previously unoccupied LL's. The restored chiral
symmetry region consists of several phases with different number
of LL's occupied, which are separated by the so-called Van Alphen
De Haas transitions, whose form is in the absence of vector mesons
approximately $\mu_c= \sqrt{2 k |q_f| B}$ (being this condition
exact for the chiral case). As a result of the different quark
electric charges, one up transitions is found every two down
transitions when a phase diagram is traversed in the magnetic
field direction at high enough fixed $\mu$. In all phase diagrams,
a ``main transition'' is found, which separates the vacuum phase
from the phases with populated LL's and at fixed $eB$ it is the
one with the lowest possible chemical potential (e.g.: lower black
line in bottom left diagram in Fig.~\ref{fig1}). In some cases,
another main transition within the populated phase exists. It
connects phases with partially restored symmetry and low LL
population to the fully restored symmetry phases, where the
Van-Alphen De Haas transitions are present and LL level population
may be much higher. All of this gives rise to a potentially
complex phase diagram, whose precise form depends on the parameter
set and magnetic field (The $B = 0$ case is simple yet depending
on the parameter set there can be a few differences). The lower
main transition usually shows, for moderate magnetic fields, a
decrease of $\mu_c$ when $eB$ increases that is sometimes called
magnetic anticatalysis~\cite{Preis:2010cq} (even though the name
is more generally used to refer to the decrease of the quark
condensates as $eB$ increases). For higher values of the magnetic
field this tendency is reverted, and this gives rise to a
characteristic curve in the main transition line to which we will
refer as the ``IMC well''. It is also worth noting that some of
the other transitions in the phase diagrams also exhibit an
IMC-like behavior.

Now we will discuss how the phase diagrams are modified as the
parameter $c_s$ is varied. The upper two panels correspond to the
$c_s=0$ case (no flavor mixing), which was previously studied in
Ref.~\cite{Allen:2013lda}. In a sense, each flavor will have its
own independent phase diagram because the gap equations are
decoupled. Here, each line corresponds to a transition where LL
population of a single quark flavor occurs (red lines for down
quarks, blue lines for up quarks). Both flavors will coincide for
$eB = 0$ where SU(2) symmetry is recovered and behave
differently as $eB$ increases, due to their different electric
charges. Since this is the only difference between the equations
for both flavors and since it only appears in the product $q_fB$,
the down flavor phase diagram may be obtained from the other one
through the replacement $q_uB = (2q_d)B = q_d(2B)$ which amounts
to a stretching of the up flavor phase diagram along the $eB$
axis. In other words, since the down quark has a smaller coupling
to the field than the up quark, it will require a magnetic field
twice as large to replicate the effect on an up quark. As a
consequence of this, the IMC wells are shifted with respect to
each other, so there will be a large well-distinguished region
where up quarks exist in the lowest LL (LLL) while down quarks are
in vacuum (0B), and another region where the opposite occurs (B0).

For finite $c_s$ (second row onwards in Fig.~\ref{fig1}), the
coupling between flavors creates a complex pattern, where
transitions move closer together to the point of  coalescing in
some regions, that is, the LL population changes simultaneously
for both flavors (these are represented by black lines). We can
see that already for $c_s=0.03$, the transitions in both parameter
sets occur together for low magnetic fields, and then separate for
$eB = 0.02$~GeV$^2 $ in Set~1 and $eB = 0.1$~GeV$^2$ in Set~2.
Note that for $c_s=0$, the transitions for both flavors cross
around $eB=0.12$~GeV$^2$ (for both sets). When flavor mixing is
introduced, this crossing point transforms into a line, that is,
both transitions occur together once again in an interval of
magnetic field values. When $c_s=0.03$, they separate at $eB =
0.4$~GeV$^2$, which is intuitive since a higher magnetic field
will further break SU(2) flavor symmetry. For $c_s = 0.2$ this
separation is no longer seen in the diagrams but it can be guessed
that it does occur beyond $eB=1$~GeV$^2$, and that for any value
of $c_s < 0.5$ there will always be a large enough magnetic field that
will cause the $u$ and $d$ main transitions to separate.

\begin{figure}
\begin{center}
\includegraphics[width=0.4\linewidth,angle=-90]{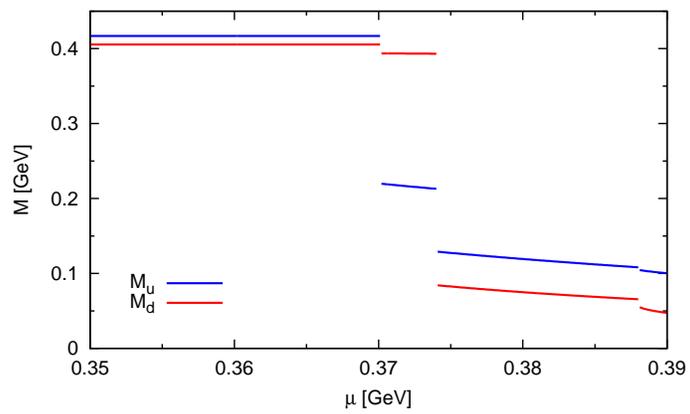}
\end{center}
\caption{\label{fig1b}(Color online) Dressed masses for both
flavors for Set~2, $c_s=0.03$, for $eB=0.11$~GeV$^2$.}
\end{figure}

For a better understanding of the physical meaning of the transition lines,
we present in Fig.~\ref{fig1b} the dressed masses for both flavors for
Set~2, $c_s = 0.03$, for $eB = 0.11$~GeV$^2$. When the first
discontinuity is encountered, at $\mu = 370$~MeV, $M_u$ jumps to
half its value and its LLL becomes populated. On the other hand,
the down flavor remains in vacuum. Actually, its mass presents a
small discontinuity caused by the weak coupling to the up quark
which is not to be interpreted as a down transition. The down
quark LLL is occupied at $\mu = 374$~MeV. This is precisely the
kind of behavior that generates a rich phase diagram for low $c_s$
values. The difference in masses is understood in terms of the
effect of magnetic catalysis. Since the flavors have different
charges, they couple with different intensities to the magnetic
field, so the up quark will have a larger mass in the vacuum phase
and a lower one in the populated phases, which is consistent with
the fact that mass increases with magnetic field in the vacuum
phase and decreases in the populated phase.

The behavior of the crossovers as $c_s$ is varied from $0$ to
$0.5$ is interesting to note. For $c_s = 0$, in Set~1, there is
one crossover for each flavor. When $c_s$ is increased, the up
crossover is slightly shifted to the left when crossing from the
$00$ to the $01$ phase, acquiring a small discontinuity. In turn,
a down crossover starts to appear from the left in the latter
phase. As $c_s$ is increased, these two crossovers move towards
each other, while the up crossover in $00$ shifts to the right,
also moving towards the down crossover originally existing in that
same phase. The $c_s = 0.5$ limit, in which down and up crossovers
have joined, is achieved very slowly, being the crossovers still
separated for $c_s \simeq 0.4995$.

The first order transitions for both flavors already occur
together in the whole studied region for $c_s = 0.2$, so the
qualitative behavior is very similar to the full flavor mixing
case $c_s = 0.5$. In fact, the model tends to full mixing quite
quickly and only for $c_s< 0.1$ are relevant mixture effects (or
more precisely, the absence of it) actually seen. VA-dH lines are
brought together for a particularly small amount of mixing,
already coinciding for $c_s = 0.03$, while crossover transitions,
on the other hand, tend much more slowly to the $c_s = 0.5$
behavior.

\subsection{Effect of the vector interactions}

In this section we analyze the effect of vector interaction terms.
As discussed in the previous section, for $c_s \gtrsim 0.1$ the
phase diagrams do not present qualitative variations. Here,
therefore, we consider $c_s=0.2$ which also is in the range of
realistic values suggested in Ref.~\cite{Frank:2003ve}. Although
the results to be shown below correspond to $c_v=0$, our studies
show that only small quantitative differences occur when varying
this parameter from $0$ to $1/2$.

\begin{figure}
\begin{center}
\includegraphics[width=0.8\linewidth,angle=0]{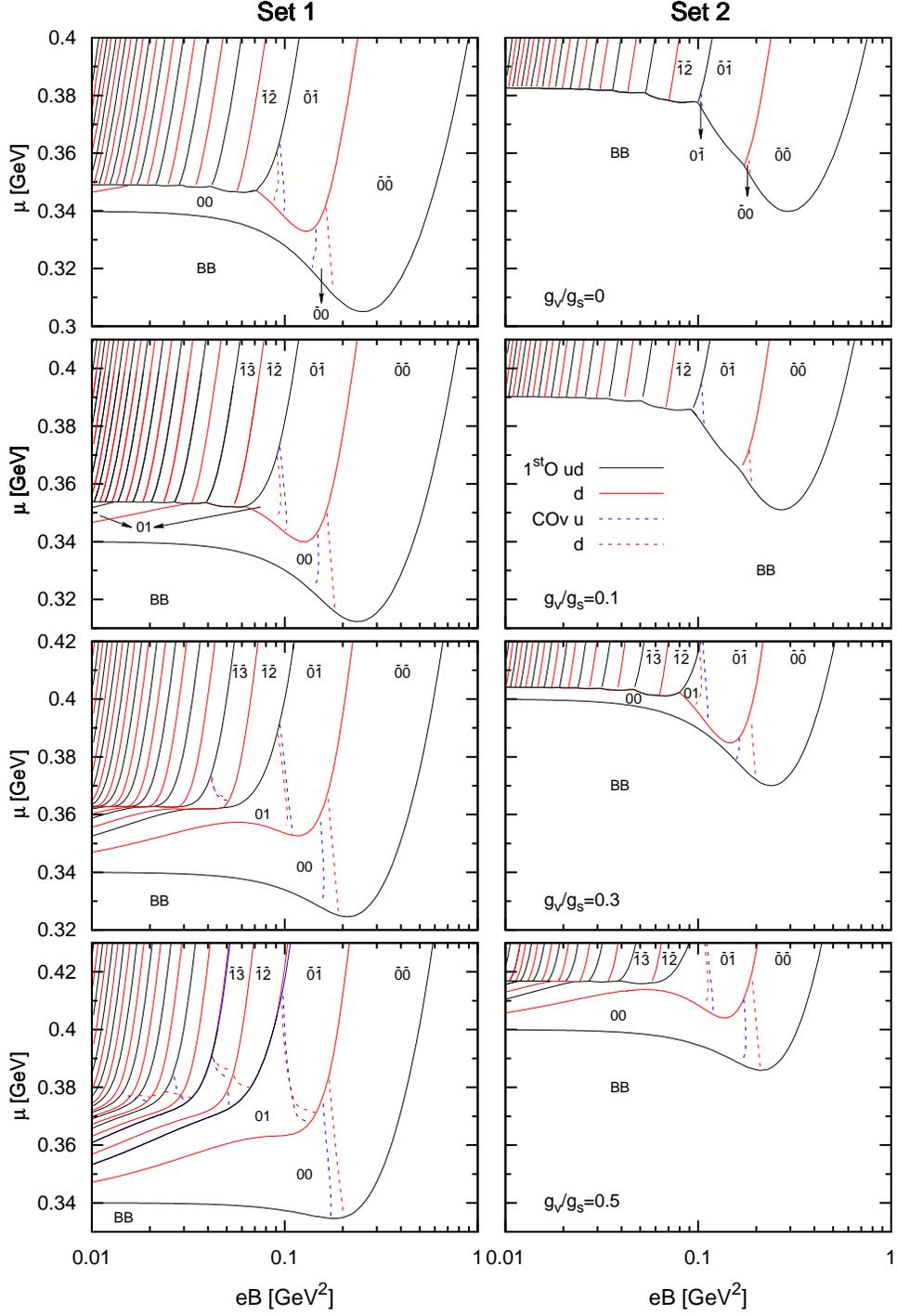}
\end{center}
\caption{\label{fig2}(Color Online) Phase diagrams in the $eB-\mu$
plane for different values of $g_v/g_s$. Different phases are
denoted as in Fig.~\ref{fig1}}
\end{figure}

\begin{figure}
\begin{center}
\includegraphics[width=0.4\linewidth,angle=-90]{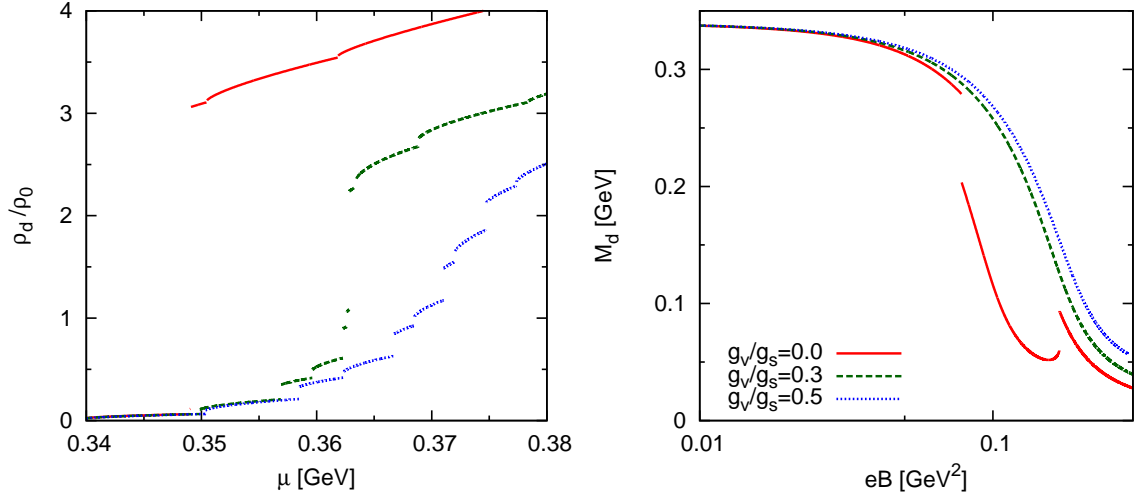}
\end{center}
\caption{\label{fig2b}(Color online) Quark density over nuclear
matter density as function of the chemical potential for
$eB=0.016$~GeV$^2$ (left) and current mass as function of magnetic
field for $\mu=0.345$~GeV (right) for different values of
$g_v/g_s$. Results correspond to $d$ flavor and were obtained with
parameter Set~1.}
\end{figure}

In Fig.~\ref{fig2} we present a series of phase diagrams obtained
for different values of the ratio $g_v/g_s$. For Set~1 we observe
that as $g_v/g_s$ increases the two main transitions separate and
several new transitions appear in between, in the low $eB$ region
of the diagram. These are partially restored symmetry regions,
where quark mass acquires an intermediate value. For Set~2 there
is a unique main transition for $g_v=0$, but already for $g_v/g_s
\simeq 0.1$ it will split into two, leaving a $00$ phase in
between which was not present before. For larger values of
$g_v/g_s$ the behavior is similar to that of Set~1. The existence
of new transitions at low $eB$ as $g_v/g_s$ increases can also be
appreciated in the left panel of Fig.~\ref{fig2b}, where we show
$d$ quark density normalized to nuclear matter density
($\rho_0=0.17$ $fm^{-3}$) as a function of chemical potential for
$eB=0.016$~GeV in Set~1. In the absence of vector interaction, the
density jumps from close to $0$ to $3$ times nuclear matter
density, while quark population jumps from $00$ to a phase where
several LL's are occupied. As $g_v$ increases, the amount of
transitions increases too. In fact, increasing the vector
interaction coupling has the same effect as going to a parameter
set that reproduces a lower value of current mass $M_0$, described
in Ref.~\cite{Allen:2013lda}. Notice that the phase diagram for
Set~2 and $g_v/g_s=0.3$ is in this sense very similar to Set~1 and
$g_v/g_s=0$.

It is interesting to analyze the effect of the vector interactions
on the so-called inverse magnetic catalysis (IMC) mentioned in the
Introduction. We recall that the IMC is usually related to a
decrease of the critical chemical potential at intermediate values
of the magnetic fields, a phenomenon that can be clearly observed
in all the phase diagrams plotted in Figs.~\ref{fig1} and
\ref{fig2}. However, while from Fig.~\ref{fig1} we see that the
variation of the strength of flavor mixing interactions has
basically no effect on the IMC effect, the situation is different
for the vector interactions. In fact, from Fig.~\ref{fig2} we note
that if we measure the depth of the IMC well as the difference in
$\mu$ between the lowest critical chemical potential at vanishing
magnetic field and the lowest possible one in the whole diagram,
we find that this difference is reduced by an $84\%$ for Set~1 and
by a $67\%$ for Set~2 when going from $g_v=0$ to $g_v/g_s=0.5$. To
explain this feature we recall that the IMC effect can be
understood in terms of the extra cost in free energy to form a
fermion-antifermion condensate at finite $\mu$,
$\Omega_{ext}$~\cite{Preis:2010cq}. In the absence of vector
interactions this extra cost basically originates in the LLL
contribution to the medium term. Considering the chiral limit for
simplicity, this contribution can be shown to be proportional to
$B\mu^2$ for symmetric matter, and it tends to decrease the
difference in free energy between the vacuum phase and the finite
density phase. As it is clear from Eqs.~(\ref{PmuB}, \ref{Omepot})
the presence of the vector interactions introduces some
modifications in $\Omega_{ext}$. In the case of the medium term,
they imply the replacement $\mu_f \rightarrow \tilde \mu_f = \mu_f
- \omega_f$. Moreover, there is a new contribution coming from
$\Omega_{pot}$. Thus, assuming as above that there is LLL
dominance and that quarks are massless in the chirally restored
phase, the extra cost for symmetric matter is
\begin{eqnarray}
\Omega_{ext} = \frac{N_c B}{4 \pi^2} \sum_f |q_f| (\mu-\omega)^2 +
\frac{\omega^2}{4 g_v} \label{omext}
\end{eqnarray}
where, for simplicity, we have assumed $c_v=1/2$. The
generalization for arbitrary values of $c_v$ is straightforward
and, in addition, the numerical dependence on $c_v$ of the
estimates to be given below turns out to be negligible. Of course,
$\omega$ should satisfy the associated gap equation which follows
from Eq.~(\ref{gapeqs}) in the Appendix. Within the above
mentioned approximations, the solution of this equation is
\begin{equation}
\omega = \left[ 1 + \frac{\pi^2}{N_c g_v \sum_f |q_f|
B}\right]^{-1} \mu
\end{equation}
Replacing in Eq.~(\ref{omext}) we get that $\Omega_{ext}$ can be
expressed in a form similar to that obtained in the absence of
vector interactions, but where $\mu$ has to multiplied by a factor
$1/\sqrt{ 1 + N_c g_v eB /\pi^2}$. Note that the actual values of
the $u-$ and $d-$quark charge have been already used to obtain
this factor. Thus, in the region $eB\sim 0.2$~GeV$^2$ around which
this expression is approximately valid we expect
\begin{eqnarray}
\frac{ \mu_c(g_v) }{ \mu_c(0) } \simeq \sqrt{ 1 + \frac{N_c}{\pi^2}\ eB\ g_v}
\end{eqnarray}
For example, for Set~2 we obtain that the lowest critical chemical potential
(which occurs at about $eB=0.24$~GeV$^2$) increases by about 10 \%
when going from $g_v=0$ to $g_v/g_s=0.5$. Together with the fact
that the lowest $\mu_c$ at vanishing magnetic field stays basically constant for
values of $g_v/g_s \gtrsim 0.1$ this explains the strong reduction of the
IMC well.

Another aspect of the reduction of the IMC phenomenon induced by
the presence of the vector interactions can be observed in the
right panel of Fig.~\ref{fig2b}. There, we plot the current quark
mass for $d$ quarks as a function of the magnetic field for Set~1,
$\mu=345$~MeV and several $g_v/g_s$ values, where the system is in
the $00$ phase at $eB=0$. As discussed in
Ref.~\cite{Allen:2013lda} in this case an actual decrease of the
mass as $eB$ increases is expected to exist. As we see, however,
such a decrease is slower for larger values of $g_v$. It should be
noticed that discontinuities appearing for $g_v/g_s=0$ corresponds
to the $d$ quark transition from the phase $00$ to $\bar 0 \bar 1$
and back.

% This values are summarized in table [...].
%
% \begin{tabular}{|c|c|c|c|c|}
% \hline
% $g_v/g_s=$ & $0$ & $0.1$ & $0.3$ & $0.5$\\
% \hline
% Set 2: & $42.8$ & $39.21$ & $29.85$ &  $14.04$\\
% Set 1: & $34.3$ &  $27.62$ &  $13.37$ &  $5.4$\\
% \hline
% \end{tabular}

\section{Results for stellar matter}

We now turn our attention to stellar matter, that is, matter where
$\beta$-equilibrium and charge neutrality are imposed. In this
case, electrons and muons are introduced into the system so that
the thermodynamical potential receives an extra
contribution~\cite{Menezes:2008qt}
\begin{eqnarray}
\Omega^{lep} &=& \sum_{l=e,\mu} \sum_{\nu=0}^{\nu^{max}_l}
\frac{|q_{l}|B\ \alpha_\nu}{4\pi^{2}}  \left[
\mu_l\sqrt{\mu^{2}-s_{l}(\nu,B)^{2}}
%\right.\nonumber\\
%& & \left. \qquad \qquad \qquad
-s_{l}(\nu,B)^{2}\ln\left(
\frac{\mu_l+\sqrt{\mu_l^{2}-s_{l}(\nu,B)^{2}}}{s_{l}(\nu,B)}
\right) \right] \label{omelep}
\end{eqnarray}
where $\nu^{max}_l = \mbox{Int}[(\mu_l^2 - m^2_l)/(2 |q_l| B)]$
and $s_{l}(\nu,B)=\sqrt{m_{l}^2+2|q_{l}|B\nu}$. We take $m_e=
0.511$~MeV and $m_\mu =  105.66$~MeV.

The $\beta$-equilibrium and charge neutrality conditions read
\begin{eqnarray}
\mu_d = \mu_u + \mu_e \qquad , \qquad \mu_e = \mu_\mu
\label{neutr}
\end{eqnarray}
and
\begin{eqnarray}
\rho_e + \rho_\mu = \frac{1}{3}
\left( 2 \rho_u - \rho_d\right) \ ,
\label{charge}
\end{eqnarray}
respectively. The lepton densities appearing in the last equation
can be easily obtained from the derivatives of the total
thermodynamical potential with respect to the corresponding
chemical potentials.

Following the discussions in the previous section only results for $c_s=0.2$
will be presented. If charge neutrality is imposed on our system,
there will be a fixed relation between the densities of the up and
down quarks, which will be necessarily different unless they are
both zero. So, even though the value of $c_s=0.2$ was close enough
to the full mixing case according to what was established in
previous sections, charge neutrality will cause the flavors to
behave differently among themselves.

\begin{figure}
\begin{center}
\includegraphics[width=0.8\linewidth,angle=0]{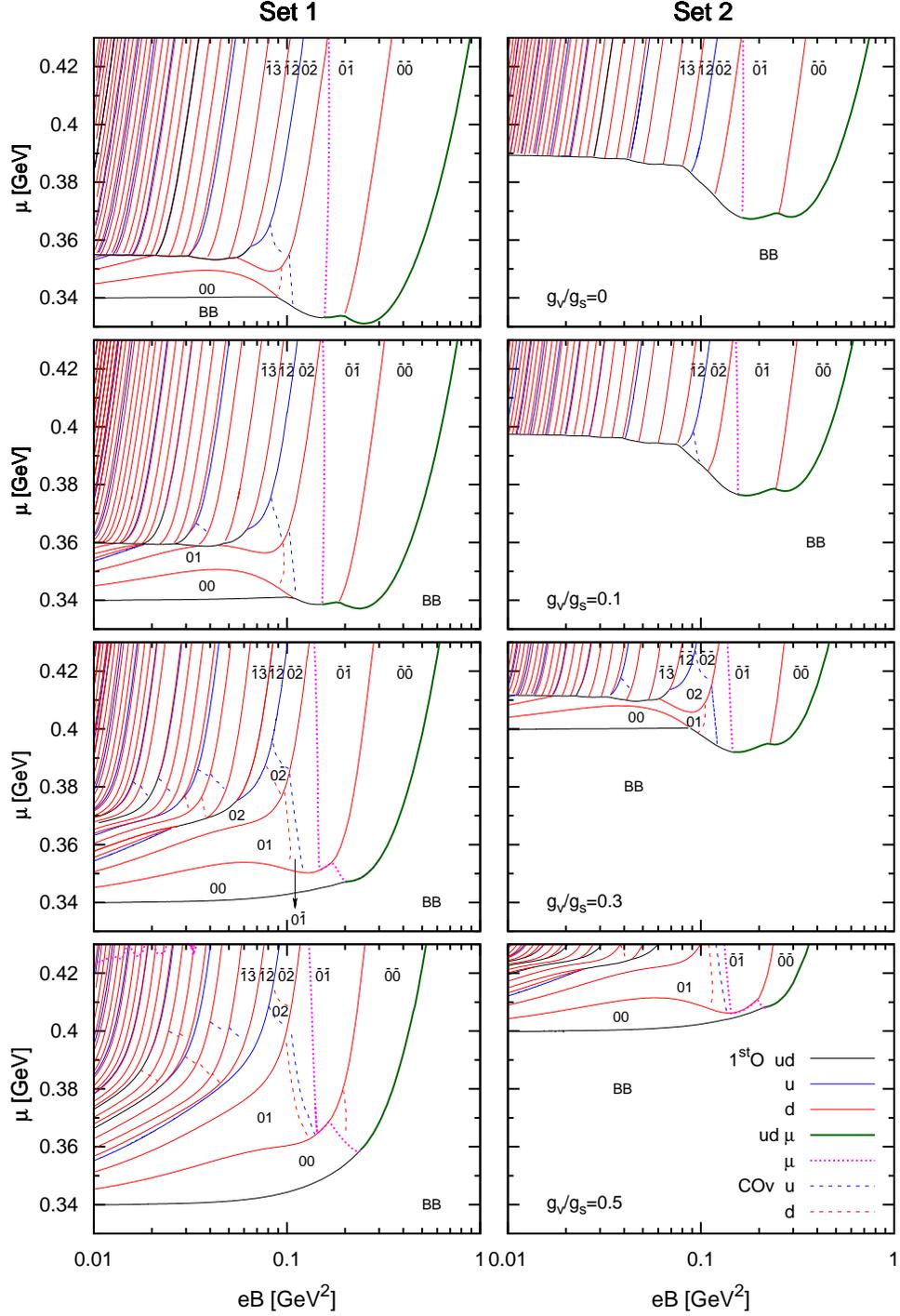}
\end{center}
\caption{\label{fig3}(Color online) Phase diagrams in the $eB-\mu$
plane for stellar matter and different values of $g_v/g_s$. Different
phases are denoted as in Fig.~\ref{fig1}. The pink dotted line represents
muon transition from vacuum to LLL.}
\end{figure}

The phase diagrams in the $\mu-B$ plane for both parameter sets
and for increasing values of $g_v$ are plotted in Fig.~\ref{fig3}.
The chemical potential $\mu$ on the horizontal axis is now the
quark number chemical potential, in terms of which the flavor
chemical potentials read $\mu_f = \mu - q_f \mu_e$ when
Eqs.(\ref{neutr}) are used. The introduction of stellar matter
conditions has a few effects similar to those of vector
interaction, in the sense that diagrams become similar to the ones
corresponding to lower $M_0$ sets: main transitions separate for
low $eB$ and several transitions appear in the region between
them. The magnetic anticatalysis effect is reduced also: The depth
of the anticatalysis well, as defined in the previous section, is
reduced from $35$~MeV to $9$~MeV in Set~1 and from $43$~MeV to
$22$~MeV in Set~2. As discussed in Ref.~\cite{Grunfeld:2014qfa}
this can be understood by generalizing to stellar matter the
discussion given in Sec.IIIb. We see that again the minimum
critical chemical potential occurs at $eB \sim 0.2$~GeV$^2$.
Around that value, and in the absence of vector interactions, the
extra cost in free energy to form a fermion-antifermion pair at
finite $\mu$ is in this case proportional to $B \bar \mu^2$ with
$\bar \mu^2 = \sum_f |q_f| \mu^2_f + \mu^2_e/3$. Here, a generally
small muonic contribution has been neglected. Using the relations
obtained from the $\beta$ equilibrium conditions, $\mu_f = \mu -
q_f \mu_e$, we get $\bar \mu = \mu (1 - 2x/3 + 2x^2/3)^{1/2}$,
where $x=\mu_e/\mu$. Note that the minus sign in the (dominant)
linear term follows from the fact that $|q_u|=2 |q_d|$. The
relevant value of $x$ follows from the neutrality condition
Eq.~(\ref{charge}). Assuming as before that in the chirally
restored phase we are dealing with massless quarks one obtains $x
\simeq 0.38$ for $\mu \simeq 350$~MeV. Using this result we get
$\bar \mu \simeq 0.92 \ \mu$. This implies that the extra cost in
free energy is smaller than that required in the symmetric matter
case for the same value of $eB$ and $\mu$. Consequently, for a
given $eB$, one needs a larger value of the chemical potential to
induce the phase transition. In fact, we have
$\mu_c^{st}/\mu_c^{sym} \simeq 1.09 $ a value which is in good
agreement with our full numerical results. In principle to
determine the change in the IMC well we should also estimate the
modification of $\mu_c$ at $eB=0$. As shown in
Ref.~\cite{Grunfeld:2014qfa} this value also increases when
stellar conditions are imposed. However, such an increase is
several times smaller than the one at $eB \sim 0.2$~GeV$^2$
leading to a quenching of the IMC effect. As discussed in
Sec.IIIb, the introduction of vector interaction will further
enhance these effects. Consequently we observe that anticatalysis
has completely disappeared for $g_v=0.5$.

The VA-dH transitions for different flavors also acquire
relatively independent behaviors when charge neutrality is
introduced. In particular, some down and up transitions occur
simultaneously as we follow them upwards along the phase diagram
and then separate visibly at a given chemical potential. This is
most clearly seen at intermediate $eB\,(\simeq 0.1$~GeV$^2$) for
the transitions separating the $02$, $12$ and $13$ phases. The
relationship for critical chemical potentials $\mu_c= \sqrt{2 k
|q_f| B}$, which occurred for symmetric matter now holds for each
flavor chemical potential separately, where these two are
different between themselves and related through
Eq.~(\ref{neutr}). As a result of this, we see that in
Fig.~\ref{fig3}, which is instead plotted in terms of quark
chemical potential, we roughly have one up transition every four
down transitions, whereas in symmetric matter there was one up
transition every two down transitions. In particular, for high
$g_v$, the transitions have a tendency to clump up in groups of
three (two down transitions and one up) with an intermediate down
transition which is well separated from this group.

 The transitions corresponding to the population of lepton
Landau Levels are also included in Fig.~\ref{fig3}. In all cases,
it was seen that electron's transition from vacuum to LLL occurs
in the lower main transition, that is, quarks and electrons occupy
their LLL simultaneously. This is understandable in that electrons
have a very small, almost negligible mass, hence they will
populate their LLL as soon as the associated chemical potential
becomes finite. On the other hand, muon behavior is more complex.
Since they have a larger mass than electrons it is expected that
they will require a larger chemical potential to acquire a finite
population. This is what actually happens in all our diagrams for
low magnetic fields, where values larger than $430$~MeV are needed
for this, hence not appearing in the plotted region. However, at
intermediate values magnetic field values ($0.1<eB<0.2$~GeV$^2$),
this transition drops suddenly until it joins the lower main
transition of the quarks. This happens because $\mu_e$ increases
with magnetic field, allowing for muon population at lower quark
chemical potential. For higher $eB$ values, all leptons and quark
transitions occur together. The occupation of the first Landau
Level of either lepton species requires much higher chemical
potentials. This happens because leptons have a larger charge than
quarks, leading to Landau Levels with larger energy, and because
$\mu_e$ is only a slowly increasing function of $\mu$. The value
needed for the first electron LL to be occupied, which is around
$250$~MeV, is therefore only achieved near $\mu=900$~MeV.  In some
phase diagrams, the muon transition is seen to affect the behavior
of the crossovers. This occurs because the $\mu_e$ is coupled to
the order parameters, so a muon transition, with its corresponding
jump in $\mu_e$, will affect the values of the quark masses and
its associated susceptibilities. For $g_v/g_s=0.5$ in Set~1, for
example, the down crossover separating the $00$ and $\bar{0}
\bar{0}$ phases is smeared out near the muon transition. In other
diagrams, like the one for $g_v/g_s=0.3$, the muon transition has
completely absorbed the quark crossover.

\begin{figure}
\begin{center}
\includegraphics[width=0.4\linewidth,angle=-90]{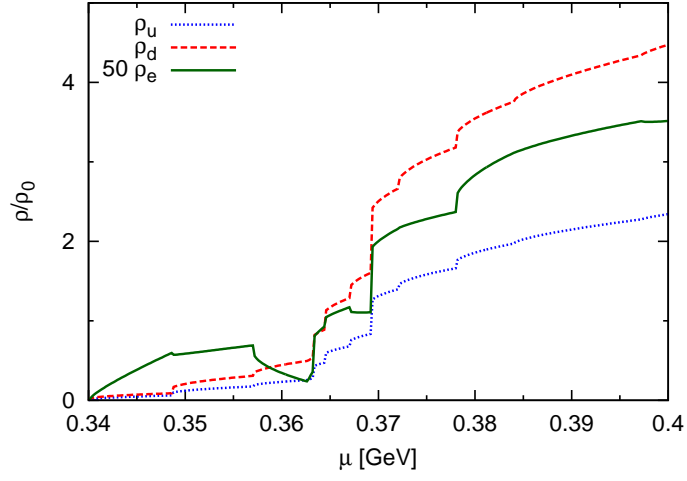}
\end{center}
\caption{\label{fig3b}(Color online)
Quark and electronic densities over nuclear
matter density as functions of the chemical potential for
$eB=0.02$~GeV$^2$. Results were obtained with
$g_v/g_s=0.3$ and parameter Set~1.}
\end{figure}

In Fig.~\ref{fig3b} we plot the normalized lepton densities
together with those of the quarks as functions of the quark number
chemical potential, for $eB=0.02$~GeV$^2$, $g_v/g_s=0.3$ and
parameters of Set~1. All three densities become non zero
simultaneously at $\mu=340$~MeV, where they all occupy their
corresponding LLL. However, while quark density grows monotonously
after the first transition, it is seen that when the next quark
transition is encountered $\rho_e$ decreases, and jumps to an even
lower value in the transition that follows. Recalling that these
correspond to transitions of $d$ quarks and that the charge
neutrality condition imposed is $\rho_e+\rho_{\mu}=(2\rho_u-\rho_d)/3$,
it can be understood that the electron density decreases whenever
there is a $d$-transition and subsequent growth of $\rho_d$ (recall
that $\rho_{\mu}=0$ in this range of chemical potential), without relevant changes in
$\rho_u$. This is the case of this range of chemical potential and
this value of magnetic field.

\begin{figure}
\begin{center}
\includegraphics[width=0.7\linewidth,angle=0]{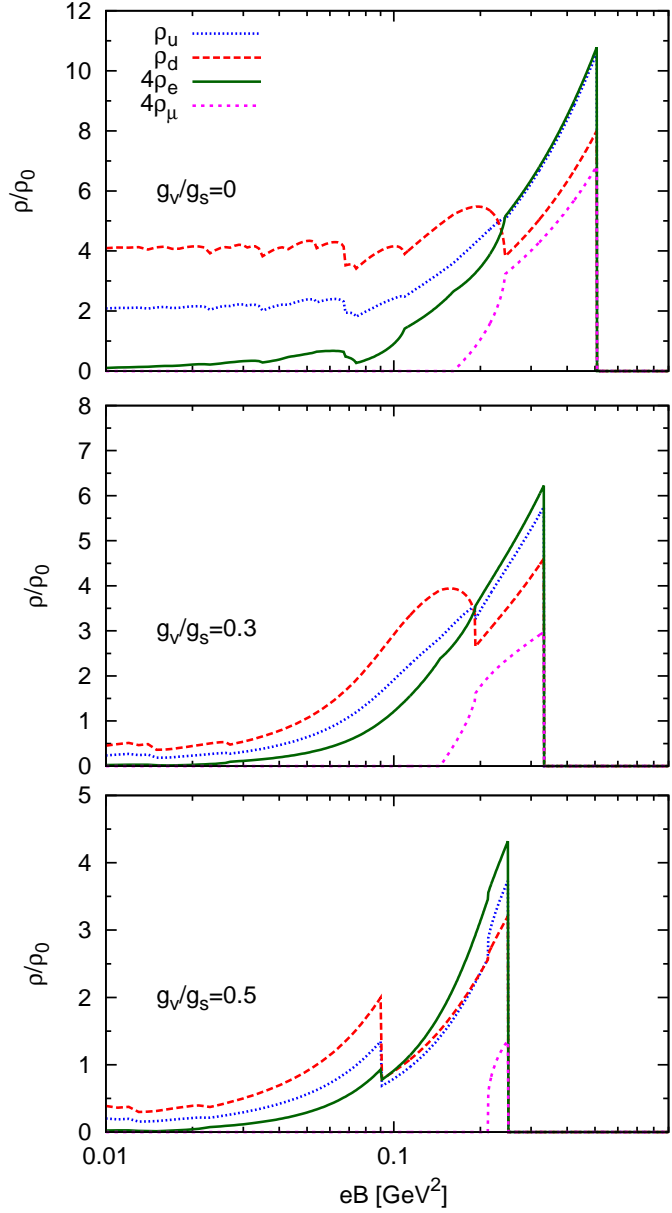}
\end{center}
\caption{\label{fig3c}(Color online)
Quark and lepton densities over nuclear
matter density as function of the magnetic field for
$\mu=0.36$~GeV for different values of
$g_v/g_s$. Results were obtained with
parameter Set~1.}
\end{figure}

In Fig.~\ref{fig3c} we show quark and lepton densities as
functions of the magnetic field for $\mu=0.36$~GeV, for different
strengths of the vector coupling. Results correspond again to
Set~1. At low magnetic field, discontinuities in the quark
densities, associated with the VA-dH transitions, are always
encountered. For the cases with $g_v=0$ and $0.3$ a region is
found in which $\rho_d$ changes its monotony within a C-type
phase. This is because quarks can be found in this phase for
magnetic fields high enough so the mass has started to increase
with $eB$, and is associated to the region of the phase diagram
above the IMC well. This behavior is not observed for
$g_v/g_s=0.5$ where the IMC effect has disappeared. For $eB\sim
0.2$~GeV$^2$, $d$ quarks are already in a phase with a low $k$ and
$u$ quarks with $k=0$, all densities increase and it is
interesting to notice how slopes change whenever a $d$ transition
is encountered. When $d$ quarks finally reach the $0$ phase all
densities increase until they drop to zero when the last
transition to vacuum is encountered.

\section{Conclusions}

We have investigated how the phase diagram of cold strongly
interacting matter is modified in the presence of intense magnetic
fields in the context of NJL-type models which include flavor
mixing and vector interaction. The whole range of possible flavor
mixing values was swept through, going from the situation in which
the two flavors are completely decoupled, to the one in which they
are fully mixed. For low mixing values, a complex phase diagram is
generated, but already for $c_s\simeq 0.1$, phase diagrams display
a behavior that is qualitatively very similar to the full mixing
case. Since SU(3) estimates suggest a value of $c_s\simeq 0.2$, it
can be concluded that the realistic flavor mixing range is similar
in behavior to the full mixing case. In what followed, vector
interactions and stellar matter conditions were introduced.
% separately, and then we studied the joint effect of these two.
The most notable effect observed is that they attenuate the
Inverse Magnetic Catalysis phenomenon, to the point that it
completely disappears when both effects are jointly taken into
account. It was also found that introducing vector interaction
causes the two main transitions to separate and additional phases
to appear between these two, effects that are similar to those
occurring when changing to a parameter set that fits to a smaller
dressed mass. In the stellar matter case, the behavior of the
leptons was also studied. It was found that while electron LLL
becomes populated simultaneously with quarks, the muon transition
presents a more complicated dependence with the magnetic field.
Namely, muon LLL requires a very high chemical potential to become
populated at low $eB$, however, it joins the main quark transition
together with electrons for high enough magnetic field.

\begin{acknowledgments}
This work was partially supported by CONICET (Argentina) under
grant PIP 00682 and by ANPCyT (Argentina) under grant
PICT-2011-0113.
\end{acknowledgments}

\section*{APPENDIX}

The explicit form of the gap equations is
\begin{eqnarray}
&&\phi_u + \frac{ (1 - c_s) \sigma_u - c_s \sigma_d}{4g_s (1-2c_s)}=0
\qquad ; \qquad
\rho_u - \frac{ (1 - 2 c_v) \omega_u - c_v \omega_d}{4g_v (1-2c_v)}=0
\nonumber \\
&&\phi_d + \frac{ (1 - c_s) \sigma_d - c_s \sigma_u}{4g_s (1-2c_s)}=0
\qquad ; \qquad
\rho_d - \frac{ (1 - 2 c_v) \omega_d - c_v \omega_u}{4g_v (1-2c_v)}=0
\label{gapeqs}
\end{eqnarray}
where $\rho_f$ in the quark density for each flavor
\begin{eqnarray}
\rho_f = \frac{N_c}{2 \pi^2} |q_f| B \sum_{\nu=0}^{\nu^{max}_f} \alpha_\nu \sqrt{\tilde\mu_f^2-s_{f}(\nu,B)^{2}}
\end{eqnarray}
and $\phi_f$ can be written as the sum of three terms
\begin{eqnarray}
\phi_{f}^{\mathrm{vac}} &=& -\frac{N_c  M_{f}}{2\pi^{2}} \left[\Lambda \sqrt{\Lambda^2 + M_f^2}-{M_{f}^{2}}\ln \left ( \frac{\Lambda+ \sqrt{\Lambda^2 + M_f^2}}{{M_{f} }} \right ) \right ] \mbox{,}
\nonumber \\[2.mm]
\phi_{f}^{\mathrm{mag}} &=& -\frac{N_{c} M_{f}|q_{f}| B}{2\pi^{2}}\left [ \ln \Gamma(x_{f}) -\frac{1}{2} \ln (2\pi) \right .
+ \left . x_{f} -\frac{1}{2} \left ( 2 x_{f}-1 \right )\ln (x_{f}) \right ] \mbox{,}
\nonumber \\[2.mm]
\phi_{f}^{\mathrm{med}} &=& - \frac{N_{c}}{2\pi^{2}} \ M_f \ |q_{f}|B \sum_{\nu=0}^{\nu^{max}_f} \alpha_\nu \
 \ln\left[ \frac{\tilde\mu_f+\sqrt{\tilde\mu_f^{2}-s_{f}(\nu,B)^{2}}}{s_{f}(\nu,B)} \right] \mbox{,}
\label{conds}
\end{eqnarray}

\end{document}